\documentclass[prl,superscriptaddress,floatfix,twocolumn]{revtex4}

\usepackage{graphicx}
\usepackage{psfrag}
\usepackage{amsmath}
\usepackage{natbib}

\begin{document}
\newtheorem{theorem}{Theorem}
\newtheorem{acknowledgement}[theorem]{Acknowledgment}
\newtheorem{algorithm}[theorem]{Algorithm}
\newtheorem{axiom}[theorem]{Axiom}
\newtheorem{claim}[theorem]{Claim}
\newtheorem{conclusion}[theorem]{Conclusion}
\newtheorem{condition}[theorem]{Condition}
\newtheorem{conjecture}[theorem]{Conjecture}
\newtheorem{corollary}[theorem]{Corollary}
\newtheorem{criterion}[theorem]{Criterion}
\newtheorem{definition}[theorem]{Definition}
\newtheorem{example}[theorem]{Example}
\newtheorem{exercise}[theorem]{Exercise}
\newtheorem{lemma}[theorem]{Lemma}
\newtheorem{notation}[theorem]{Notation}
\newtheorem{problem}[theorem]{Problem}
\newtheorem{proposition}[theorem]{Proposition}
\newtheorem{remark}[theorem]{Remark}
\newtheorem{solution}[theorem]{Solution}
\newtheorem{summary}[theorem]{Summary}    
\def\r{{\bf{r}}}
\def\i{{\bf{i}}}
\def\j{{\bf{j}}}
\def\m{{\bf{m}}}
\def\k{{\bf{k}}}
\def\h{{\bf{h}}}
\def\kt{{\tilde{\k}}}
\def\qt{{\tilde{\q}}}
\def\mt{{\hat{t}}}
\def\mG{{\hat{G}}}
\def\mg{{\hat{g}}}
\def\mGa{{\hat{\Gamma}}}
\def\mS{{\hat{\Sigma}}}
\def\mT{{\hat{T}}}
\def\K{{\bf{K}}}
\def\P{{\bf{P}}}
\def\q{{\bf{q}}}
\def\Q{{\bf{Q}}}
\def\p{{\bf{p}}}
\def\x{{\bf{x}}}
\def\X{{\bf{X}}}
\def\Y{{\bf{Y}}}
\def\F{{\bf{F}}}
\def\R{{\bf{R}}}
\def\G{{\bf{G}}}
\def\bG{{\bar{G}}}
\def\mbG{{\hat{\bar{G}}}}
\def\M{{\bf{M}}}
\def\V{\cal V}
\def\tchi{\tilde{\chi}}
\def\tx{\tilde{\bf{x}}}
\def\tk{\tilde{\bf{k}}}
\def\tK{\tilde{\bf{K}}}
\def\tq{\tilde{\bf{q}}}
\def\tQ{\tilde{\bf{Q}}}
\def\si{\sigma}
\def\ep{\epsilon}
\def\hep{{\hat{\epsilon}}}
\def\al{\alpha}
\def\be{\beta}
\def\ep{\epsilon}
\def\bep{\bar{\epsilon}_\K}
\def\mep{\hat{\epsilon}}
\def\up{\uparrow}
\def\de{\delta}
\def\De{\Delta}
\def\up{\uparrow}
\def\dwn{\downarrow}
\def\ksi{\xi}
\def\etha{\eta}
\def\product{\prod}
\def\goto{\rightarrow}
\def\switch{\leftrightarrow}

\title{Systematic study of d-wave superconductivity in the 2D
  repulsive Hubbard model}

\author{T.A.~Maier} \affiliation{Computer Science and Mathematics
  Division, Oak Ridge National Laboratory, Oak Ridge, TN 37831}
\author{M.~Jarrell} \affiliation{Department of Physics, University of
  Cincinnati, Cincinnati, OH 45221}
\author{T.C.~Schulthess}\affiliation{Computer Science and Mathematics
  Division, Oak Ridge National Laboratory, Oak Ridge, TN 37831}
\author{P. R. C.~Kent} \affiliation{University of Tennessee, Knoxville, Tennessee 37996}
\author{J.B.~White}\affiliation{Computer Science and Mathematics
  Division, Oak Ridge National Laboratory, Oak Ridge, TN 37831}

\date{\today}

\begin{abstract} 

% Central ideas very rough draft
  The cluster size dependence of superconductivity in the conventional
  two-dimensional Hubbard model, commonly believed to describe high-temperature
  superconductors, is systematically studied using the Dynamical Cluster
  Approximation and Quantum Monte Carlo simulations as cluster solver. Due to
  the non-locality of the d-wave superconducting order parameter, the results
  on small clusters show large size and geometry effects. In large enough
  clusters, the results are independent of the cluster size and display a
  finite temperature instability to d-wave superconductivity.

\end{abstract}

\maketitle

Despite years of active research, the understanding of pairing in the
high-temperature ``cuprate'' superconductors (HTSC) remains one of the
most important outstanding problems in condensed matter physics. While
conventional superconductors are well described by the BCS theory, the
pairing mechanism in HTSC is believed to be of entirely different
nature. Strong electronic correlations play a crucial role in HTSC,
not only for superconductivity but also for their unusual normal state
behavior. Hence, models describing itinerant correlated electrons, in
particular the two-dimensional (2D) Hubbard model and its
strong-coupling limit, the 2D t-J model, were proposed to capture the
essential physics of the CuO-planes in HTSC
\cite{anderson:rvb,zhang:88}.
% With more than a thousand publications per year for the last ten
% years, these models are among the mostly studied models in condensed
% matter physics.
Despite the fact that these models are among the mostly studied models
in condensed matter physics, the question of whether they contain
enough ingredients to describe HTSC remains an unsolved problem.

%previous studies of DSC        
Many different techniques, from analytic to numerical have been applied to
study superconductivity in these models.  The Mermin-Wagner theorem
\cite{mw:66} and the rigorous results in Ref.~\cite{su:98} preclude
$d_{x^2-y^2}$ superconducting long-range order at finite temperatures in the 2D
models. Superconductivity may however exist -- as in the attractive Hubbard
model -- as topological order at finite temperatures below the
Kosterlitz-Thouless (KT) transition temperature \cite{kt:73}.  Recent
renormalization group studies indicate that the ground-state of the doped
weak-coupling 2D Hubbard model is superconducting with a $d_{x^2-y^2}$-wave
order parameter \citep{halboth:00}.  The possibility of $d_{x^2-y^2}$-wave
pairing in the 2D Hubbard and t-J models was also indicated in a number of
numerical studies of finite system size (for a review see \cite{dagotto:rmp}).
Only recent numerical calculations for the t-J model provided evidence for
pairing at $T=0$ in relatively large systems for physically relevant values of
$J/t$ \cite{sorella:tJ}.  Quantum Monte Carlo (QMC) simulations are also
employed to search for such a transition \cite{scalapino:99}.  These studies
indicate an enhancement of the pairing correlations in the $d_{x^2-y^2}$
channel with decreasing temperature.  Unfortunately the Fermion sign problem
limits these studies to temperatures too high to study a possible KT
transition.  Another difficulty of these methods arises from their strong
finite size effects, often ruling out the reliable extraction of low-energy
scales. In fact, a reliable finite-size scaling has only recently been achieved
in the negative-U model \cite{scalettar:negu}, where the relevant temperature
scales are much higher.  The available results for the positive-U model so far
have thus been inconclusive, and a treatment within a non-perturbative scheme
that goes beyond the conventional finite size techniques is clearly necessary
to resolve the controversy as to whether there exists finite temperature
superconductivity in these models.

In this Letter we use the Dynamical Cluster Approximation (DCA)
\cite{hettler:dca1} (for a review see \cite{maier:rev}) to explore the
superconducting instability in the 2D Hubbard model
\begin{equation}
  \label{eq:HM}
  H=-t \sum_{\langle ij\rangle, \sigma} c^\dagger_{i\sigma}c^{\phantom\dagger}_{j\sigma} +U\sum_i n_{i\uparrow} n_{i\downarrow}\,,
\end{equation}
where $c^{(\dagger)}_{i\sigma}$ (creates) destroys an electron with
spin $\sigma$ on site $i$, $n_{i\sigma}$ is the corresponding number
operator, $t$ the hopping amplitude between nearest neighbors $\langle
\dots \rangle$ and $U$ the on-site Coulomb repulsion.  In the DCA we
take advantage of the short length-scale of spin correlations in
optimally doped HTSC \cite{thurston:af} to map the original lattice
model onto a periodic cluster of size $N_c=L_c\times L_c$ embedded in
a self-consistent host. Thus, correlations up to a range $\xi\lesssim
L_c$ are treated accurately, while the physics on longer length-scales
is described at the mean-field level.  By increasing the cluster size,
it thus allows us to systematically interpolate between the
single-site dynamical mean-field result and the exact result while
remaining in the thermodynamic limit. We solve the cluster problem
using QMC simulations\cite{jarrell:dca3}.

We present results of large cluster calculations -- up to 26 sites --
that indicate that the 2D Hubbard model has a superconducting
instability at a finite temperature. This conclusion is reached due to
several factors: Simulations on small clusters, where d-wave order is
topologically allowed, show large finite size and geometry effects
leading to inconclusive results. However, since the average sign in
DCA QMC simulations is significantly larger than in finite-size QMC
counterparts, exploring lower temperatures and larger clusters becomes
possible.  In addition, the advent of new parallel vector machines,
such as the CRAY X1 at ORNL, improves the speed of these calculations
by more than one order of magnitude compared to conventional
architectures, making simulations on large clusters with a small
average sign feasible.  Within the limits of current computational
capability, we observe finite transition temperatures in the largest
affordable clusters. There the results are independent of cluster
size within the error bars, although we cannot preclude a further
small reduction in transition temperatures in yet larger clusters.

Previous DCA simulations with a cluster of four sites, the smallest
cluster that can capture $d_{x^2-y^2}$-wave pairing, with $U$ equal to
the bandwidth $W=8t$, show good general agreement with HTSC
\cite{maier:dca2}. In the paramagnetic state, the low-energy spin
excitations become suppressed below the crossover temperature $T^*$,
and a pseudogap opens in the density of states at the chemical
potential.  At lower temperatures, we find a finite temperature
transition to antiferromagnetic long-range order at low doping, while
at larger doping, the system displays an instability to
$d_{x^2-y^2}$-wave superconducting long-range order. This apparent
violation of the Mermin-Wagner theorem is a consequence of the small
cluster size studied (see also \cite{su:comment}). More recent results
obtained with a similar quantum cluster algorithm confirm the presence
of antiferromagnetism and superconductivity in the groundstate of the
2D Hubbard model \cite{senechal:04}.

With increasing cluster size however, the DCA progressively includes
longer-ranged fluctuations while retaining some mean-field character.
Larger clusters are thus expected to systematically drive the Ne\'{e}l
temperature to zero and hence recover the Mermin-Wagner theorem in the
infinite cluster size limit. In contrast, superconductivity may
persist as KT order even for large cluster sizes.

%How we chose our parameters?
Since the large cluster simulations presented here are at the limit of
current computational capabilities, we are restricted in our ability
to explore both the parameter space and different cluster sizes. We
choose the parameters to favor superconducting and antiferromagnetic
order.  In our study of superconductivity, we choose $U=4t=W/2$ (we
take $t$ as our unit of energy). While we observe that larger values
of $U$ yield higher transition temperatures in the 4-site cluster, the
smaller value of $U$ greatly reduces the sign problem and thus allows
us to simulate larger cluster sizes.  We focus on a doping of $10\% $,
where the pairing correlations are maximal for $U=W/2$.
% Are you sure about this??
To study antiferromagnetism, we focus on the undoped model and set
$U=8t$, where the Ne\'el temperature is highest.

\begin{figure}[b]
\centering 
\vspace*{-0.8cm}
\includegraphics*[width=2.5in]{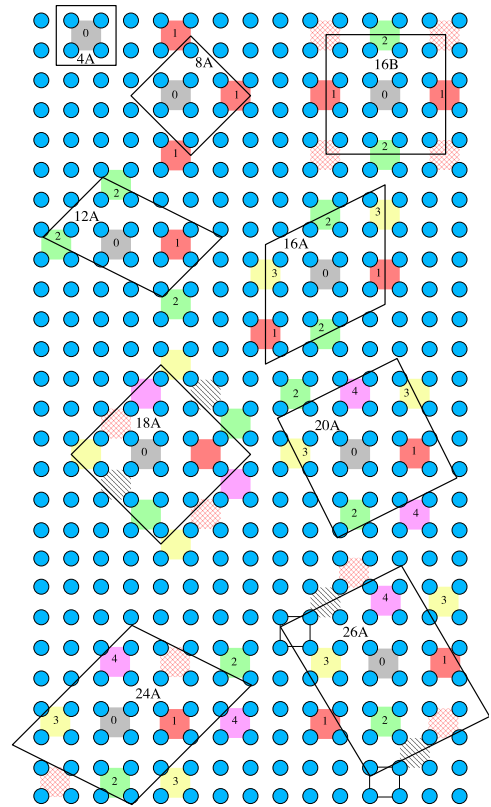} 
\vspace*{-1cm}
\caption{Cluster sizes and geometries used in our study. The shaded
  squares represent independent d-wave plaquettes within the clusters.
  In small clusters, the number of neighboring d-wave plaquettes $z_d$
  listed in table~\ref{tab:1} is smaller than four, i.e. than that of
  the infinite lattice.}
\label{fig:1}
\end{figure}

Furthermore, we have to be careful in selecting different cluster sizes and
geometries. Much can be learned from simulations of finite size systems, where
periodic boundary conditions are typically used.  Betts and Flynn
\cite{dbetts:2d} systematically studied the 2D Heisenberg model on finite size
clusters and developed a grading scheme to determine which clusters should be
used. The main qualification is the ``imperfection'' of the near-neighbor
shells: a measure of the (in)completeness of each neighbor shell compared to
the infinite lattice. In finite size scaling calculations they found that the
results for the most perfect clusters fall on a scaling curve, while the
imperfect clusters generally produce results off the curve.  Here, we employ
some of the cluster geometries proposed by Betts (see Fig.~\ref{fig:1}) to
study the antiferromagnetic transition at half filling and generalize Betts'
arguments to generate a set of clusters appropriate to study d-wave
superconductivity.

\begin{figure}[t]
  \includegraphics*[height=3.3in,angle=-90]{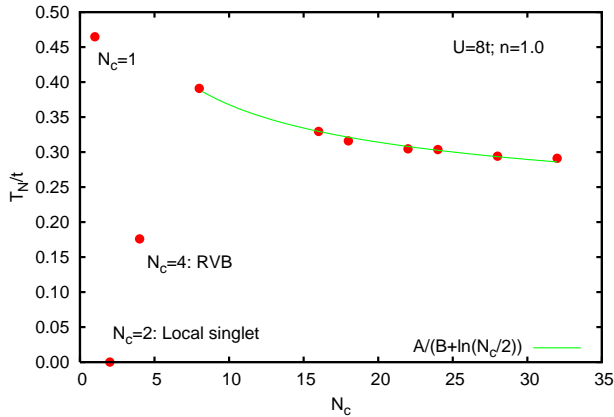}  
  \caption{Ne\'{e}l temperature at half-filling when $U=8t$ versus the
    cluster size.  $T_{\rm N}$ scales to zero in the infinite cluster
    size limit. The solid line represents a fit to the function
    $A/(B+\ln(N_c/2))$ obtained from the scaling ansatz $\xi(T_N)=L_c$.
    For $N_c=2$ a local singlet and for $N_c=4$ the RVB state suppress
    antiferromagnetism.}
  \label{fig:2}
\end{figure}

To illustrate that the DCA recovers the correct result as the cluster size
increases, we plot in Fig.~\ref{fig:2} the DCA results for the Ne\'{e}l
temperature $T_{\rm N}$ at half-filling as a function of the cluster size
$N_c$.  $T_{\rm N}$ decreases slowly with increasing cluster size $N_c$.  As
spin-correlations develop exponentially with decreasing temperature in 2D, the
$N_c>4$ data falls logarithmically with $N_c$, consistent with $T_{\rm N}=0$ in
the infinite size cluster limit.  Thus, the Mermin-Wagner theorem is recovered
for $N_c\rightarrow \infty$. The clusters with $N_c=2$ and $N_c=4$ are special
because their coordination number is reduced from four.  For $N_c=2$ the
coordination number is one and hence a local singlet is formed on the cluster
for temperatures below $J\sim t^2/U$.  In the $N_c=4$ site cluster, the
coordination is two, so fluctuations of the order parameter are overestimated
and the resonating valence bond state \cite{anderson:rvb} is stabilized.
Hence, antiferromagnetism is suppressed in these cluster sizes and their
corresponding $T_{\rm N}$ does not fall on the curve.

% in the conclusion we say up to Nc=36
We now turn to the main focus of this Letter, i.e.\ the search for a possible
KT instability to the superconducting state.  To check that the DCA formalism
is able to describe such a transition, we first tested the DCA-QMC code on the
negative $U$, i.e. attractive Hubbard model which is known to exhibit a KT
instability to an $s$-wave superconducting state \cite{scalettar:negu}. We find
that the DCA indeed produces a finite temperature s-wave instability to the KT
superconducting state. Due to the local nature of the s-wave order parameter,
the DCA results converge rather quickly with cluster size.  The DCA values for
$T_c$ agree with those recently obtained in finite size QMC simulations
\cite{scalettar:negu}. In addition, we checked that our DCA-QMC code reproduces
the results of other DMFT codes when $N_c=1$, and those of finite size QMC
codes when the coupling to the self-consistent host is turned off.

To identify a possible KT transition in the positive $U$ Hubbard model we
calculate the $d_{x^2-y^2}$-wave pair-field susceptibility $P_d$ for the
clusters $N_c=4A$, 8A, 16A, 16B, 18A, 20A, 24A and 26A. In contrast to the
s-wave order parameter in the attractive model, the d-wave order parameter is
non-local and involves four bonds or sites.  Thus, large size and geometry
effects have to be expected in small clusters. Similar to the cluster grading
scheme Betts developed for magnetic order, we can classify the different
clusters according to their quality for d-wave order. At low temperatures,
local d-wave pairs will form, but phase fluctuations of the pair wave-function
prevent the system from becoming superconducting. Since the DCA cluster has
periodic boundary conditions, each four-site d-wave plaquette has four
neighboring d-wave plaquettes. However, as illustrated in Fig.~\ref{fig:1}, in
small clusters, these are not necessarily independent and the effective
dimensionality may be reduced.

\begin{table}[b]
  \caption{Number of independent neighboring d-wave plaquettes $z_d$ and the values of $T_{c}^{\rm KT}$ and $T_c^{\rm lin}$ obtained from the Kosterlitz-Thouless and linear fits of the pair-field susceptibility in Fig.~\ref{fig:3}, respectively.}
        \begin{center}
        \begin{tabular}{ccll}
	  \hline\hline
          Cluster & $z_d$ & $T_c^{\rm KT}/t$ & $T_c^{\rm lin}/t$ \\
          \hline
           4   & 0 (MF) & 0.046   & 0.056\\
           8A  & 1      & -0.014  & -0.006 \\
           18A & 1      & -0.043  & -0.022 \\
           12A & 2      & 0.011   & 0.016  \\
           16B & 2      & 0.010   & 0.015  \\
           16A & 3      & 0.021$\pm$0.008 & 0.025$\pm$0.002 \\
           20A & 4      & 0.019   & 0.022 \\
           24A & 4      & 0.016   & 0.020 \\
           26A & 4      & 0.020   & 0.023 \\
	   \hline\hline
\end{tabular}
\end{center}
\label{tab:1}
\end{table}

Fig.~\ref{fig:1} shows the arrangement of independent d-wave
plaquettes in the clusters used in our study and their corresponding
number $z_d$ is listed in table~\ref{tab:1}. In the infinite system,
$z_{d}=4$. The $N_c=4$ cluster encloses exactly one d-wave plaquette
($z_d=0$). When a local d-wave pair forms on the cluster, the system
becomes superconducting, since no superconducting phase fluctuations
are included. Thus, the $N_c=4$ result corresponds to the mean-field
solution. In the 8A cluster, there is room for one more d-wave pair,
thus the number of independent neighboring d-wave plaquettes $z_d=1$.
Since this same neighboring plaquette is adjacent to its partner on
four sides, phase fluctuations are replicated and hence overestimated
as compared to the infinite system. The situation is similar in the
16B cluster, where only two independent (and one next-nearest
neighbor) d-wave plaquettes are found ($z_d=2$). In contrast, $z_d=3$
in the oblique 16A cluster. We thus expect d-wave pairing correlations
to be suppressed in the 16B cluster as compared to those in the 16A
cluster. With the exception of the 18A cluster, where neighboring
d-wave plaquettes share one site and thus are not independent, the
larger clusters 20A, 24A, and 26A all have $z_d=4$ and are thus
expected to show the most accurate results.  Hence, as the number of
independent neighboring d-wave plaquettes, $z_{d}$, is reduced from
four, phase fluctuations are replicated due to periodic boundary
conditions and thus overemphasized, suppressing pairing
correlations and consequently $T_{c}$.  Note that the effects of
finite size energy levels on the pairing correlations were pointed out
in QMC simulations of Hubbard ladders \cite{kuroki:1996}.
 
\begin{figure}[htbp]
  \centering
  \includegraphics*[height=3.4in,angle=-90]{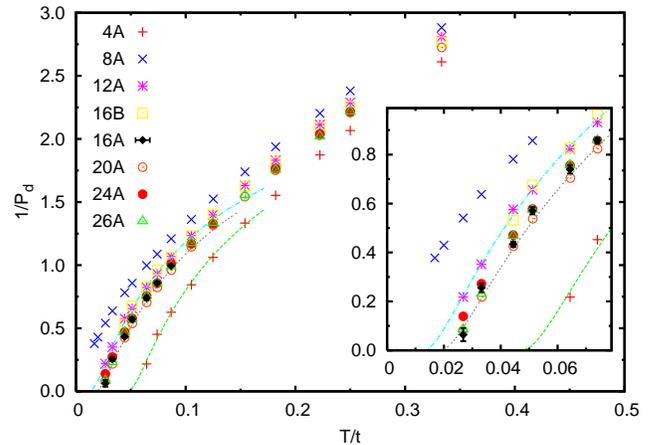}
  \caption{Inverse d-wave pair-field susceptibility as a function of
    temperature for different cluster sizes at 10\% doping. The
    continuous lines represents fits to the function
    $P_d=A\exp(2B/(T-T_c)^{0.5})$ for data with different values of
    $z_d$.  Inset: Magnified view of the low-temperature region.}
  \label{fig:3}
\end{figure}

Fig.~\ref{fig:3} shows the temperature dependence of the inverse
d-wave pair-field susceptibility, $1/P_d$, in the 10\% doped system.
% The results for the repulsive model are significantly different from
% the results for the attractive model.
Since a proper error propagation is severely hampered by storage
requirements, we obtain the error-bars shown on the 16A results from a
number of independent runs initialized with different random number
seeds. Error-bars on larger cluster results are expected to be of the
same order or larger. The results clearly substantiate the topological
arguments made above.

As noted before, the $N_c=4$ result is the mean-field result for
d-wave order and hence yields the largest pairing correlations and the
highest $T_c$. As expected, we find large finite size and geometry
effects in small clusters. When $z_d<4$, fluctuations are
overestimated and the d-wave pairing correlations are suppressed. In
the 8A cluster where $z_d=1$ we do not find a phase transition at
finite temperatures. Both the 12A and 16B cluster, for which $z_d=2$,
yield almost identical results. Pairing correlations are enhanced
compared to the 8A cluster and the pair-field susceptibility $P_d$
diverges at a finite temperature. As the cluster size is increased,
$z_d$ increases from 3 in the 16A cluster to 4 in the larger clusters,
the phase fluctuations become two-dimensional and as a result, the
pairing correlations increase further (with exception of the 18A
cluster). Within the error-bars (shown for 16A only), the results of
these clusters fall on the same curve, a clear indication that the
correlations which mediate pairing are short-ranged and do not extend
beyond the cluster size.
% Note that the larger clusters will also include pairing fluctuations
% due to next and next-next near neighbor plaquettes, which apparently
% have a very weak effect on superconductivity.

The low-temperature region can be fitted by the KT form
$P_d=A\exp(2B/(T-T_c)^{0.5})$, yielding the KT estimates for the
transition temperatures $T_c^{\rm KT}$ given in table~\ref{tab:1}. We
also list the values $T_c^{\rm lin}$ obtained from a linear fit of the
low temperature region, which is expected to yield more accurate
results due to the mean-field behavior of the DCA close to $T_c$
\cite{maier:rev}. For all clusters with $z_d\geq 3$ we find a
transition temperature $T_c\approx0.023t\pm 0.002t$ from the linear
fits.  We cannot preclude, however, the possibility of a very slow,
logarithmic cluster size dependence of the form
$T_c(N_c)=T_c(\infty)+B^2/(C+\ln (N_c)/2)^2$ where $T_c(\infty)$ is
the exact transition temperature. In this case it is possible that an
additional coupling between Hubbard planes could stabilize the
transition at finite temperatures.

In summary, we have presented DCA/QMC simulations of the 2D Hubbard
model for clusters up to $N_c=32$ sites.  Consistent with the
Mermin-Wagner theorem, the finite temperature antiferromagnetic
transition found in the $N_c=4$ simulation is systematically
suppressed with increasing cluster size. In small clusters, the
results for the d-wave pairing correlations show a large dependence on
the size and geometry of the clusters. For large enough clusters
however, the results are independent of the cluster size and display a
finite temperature instability to a d-wave superconducting phase at
$T_c\approx 0.023t$ at 10\% doping when $U=4t$.

We acknowledge useful discussions with M. Novotny, R. Scalettar, S.
Sorella, and S. R. White.  This research was enabled by computational
resources of the Center for Computational Sciences and was sponsored
by the offices of Basic Energy Sciences and Advanced Scientific
Computing Research, U.S.  Department of Energy. Oak Ridge National
Laboratory, where TM is a Eugene P.  Wigner Fellow, is managed by
UT-Battelle, LLC under Contract No. DE-AC0500OR22725. The development
of the DCA formalism and algorithm was supported by the NSF under
Grant No.  DMR-0312680 as well as through resources provided by San
Diego Supercomputer Center under NSF cooperative agreement
SCI-9619020.

\bibliography{mybib}

\end{document}